\documentclass[twocolumn,runningheads]{svjour2}
\smartqed  
\usepackage{graphicx}
\journalname{Astrophysics and Space Science}
\begin{document}

\title{H.E.S.S. Observations of LS~5039
}
\subtitle{Discovery of the 3.9 days orbital periodicity in Very High Energy $\gamma$-rays.}


\author{Mathieu de Naurois for the H.E.S.S. Collaboration.
}

\authorrunning{M. de Naurois} 

\institute{M. de Naurois \at
           Laboratoire de Physique Nucl\'eaire et de Hautes Energies\\
           4 place Jussieu,  75252 Paris Cedex 05 \\
           Tel.: +33-1-44272324\\
           Fax: +33-1-44274638\\
              \email{denauroi@in2p3.fr}           
}

\date{Received: date / Accepted: date}

\maketitle

\begin{abstract}

Recent observations of the binary system \linebreak LS~5039 with the High Energy Stereoscopic System \linebreak  (H.E.S.S.)
revealed that its Very High Energy (VHE) $\gamma$-ray emission is modulated at the 3.9 days orbital period of the system. 
The bulk of the emission is largely confined to half of the orbit, peaking around the inferior conjunction 
epoch of the compact object. The flux modulation provides the first indication of $\gamma$-ray absorption by
pair production on the intense stellar photon field. This implies that the production region size must be 
not significantly greater than the gamma-gamma photosphere size ($\sim 1\,\mathrm{AU}$), thus excluding the large scale collimated 
outflows or jets (extending out to $\sim 1000\,\mathrm{AU}$). 
A hardening of the spectrum is also observed at the same  epoch between 0.2 and a few TeV which is unexpected
under a pure absorption scenario and could rather arise from variation with phase in the maximum electron 
energy and/or the dominant VHE $\gamma$-ray production mechanism. 
This first-time observation of modulated $\gamma$-ray emission allows precise tests of the acceleration and emission models in binary systems.
\end{abstract}

\section{Introduction}
\label{intro}
In the commonly accepted paradigm, microquasars consist of a stellar mass black hole fed by a massive star. 
They can exhibit superluminous radio jets\cite{Mirabel:Nature94}, and hints for the presence of an accretion disk.
These {\it scaled down} versions of Active Galactic Nuclei (AGN) also show, due to their much lower mass, 
variability on timescales shorter by several order of magnitudes (down to minutes or even seconds) that 
can be used to constrain the accretion and ejection scenarios (e.g. \cite{Fender:1997}).

It has been suspected for a long time that these object could emit, through similar acceleration mechanisms 
as in AGN (leptonic or hadronic processes) high energy radiation up to the Very High Energy gamma-ray (VHE) 
domain, and that VHE radiation could give insight on the very central engine. 
The discovery  of VHE emission from LS~5039 by 
HESS\cite{Aharonian:HESS_LS5039} and shortly after from LS~I~+$61^\circ~ 303$ by MAGIC\cite{Albert:MAGIC_LSI} 
confirmed this longstanding issue and established VHE astronomy as a powerful diagnostic probe of these
objects.

LS~5039, identified in 1997\cite{Motch:1997} as a massive X-ray binary system with faint radio emission\cite{Marti:1997},
was resolved in 2000\cite{Paredes:2000} into bipolar mildly relativistic radio jets ($v\sim 0.2\ c$)
emanating from a central core, thus placing it into the {\it microquasar class}. 
The detection of radio and variable X-ray emission\cite{Bosch:2005} and its possible association with 
the EGRET source 3EG J1824-1514\cite{Paredes:2000} suggests the presence of multi-GeV particles
possibly accelerated in jets. LS~5039 is indeed the only object simultaneously detected in X-ray and 
radio in the field of view of the unidentified EGRET source\cite{Ribo:Thesis}.

\begin{figure}
\centering
  \includegraphics[width=0.4\textwidth]{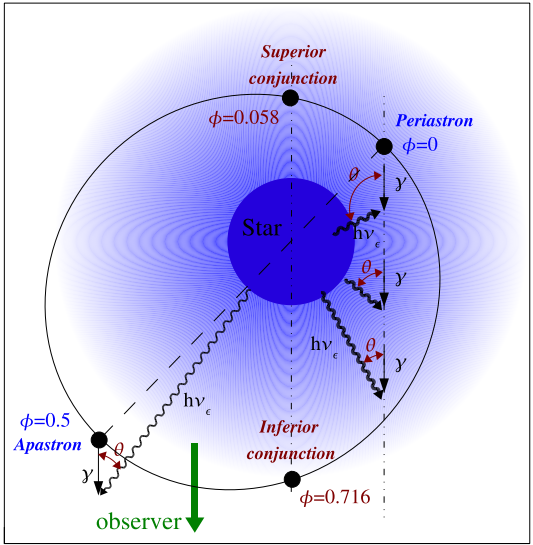}
\caption{Orbital geometry of the binary system LS~5039 viewed from above and using the orbital
parameters derived in\cite{Casares:2005}. From \cite{Aharonian:HESS_LS5039:2}. Shown are:
phases ($\phi$) of minimum ({\it periastron}) and maximum ({\it apastron}) binary separation;
epoch of superior and inferior conjunctions occurring when the compact object and the star are
aligned along the observer light-of-sight.}
\label{fig:Cartoon}       
\end{figure}

The binary system LS~5039 (Fig \ref{fig:Cartoon}) consists of a massive 06.5V star in a $\sim 3.9$ day mildly 
eccentric orbit ($e = 0.35$)\cite{Casares:2005} around a compact object whose exact nature (black hole or neutron star) is still under debate.
Under the assumption of pseudo-synchronization, Casares and collaborators\cite{Casares:2005} constrain the compact object 
mass in the black hole range ($M_X = 3.7_{-1}^{+1.3} M_\odot$) and obtain a low system inclination ($i \sim 25^\circ$),
but a neutron star at higher inclination ($i \sim 60^\circ$) might still remain possible\cite{Dubus:2006b}.


\section{H.E.S.S. Observations}

The High Energy Stereoscopic System (H.E.S.S.) is an array of four identical
Atmospheric Cherenkov Telescopes (ACT)\cite[and references therein]{Aharonian:HESS_Crab} 
located in the Southern Hemisphere (Namibia, 1800 m a.s.l.) and sensitive to
$\gamma$-rays above 100~GeV. 

LS~5039 was serendipitously detected in 2004 during the H.E.S.S. galactic scan\cite{Aharonian:HESS_LS5039}.
The 2004 observations have been followed up by a deeper observation campaign\cite{Aharonian:HESS_LS5039:2}
in 2005, leading to a total dataset of 69.2 hours of observation after quality 
selection. To optimize the coverage over the orbit, the observations were spread over more than six months,
resulting in a wide range of observation conditions. The observation zenith angles are in particular 
distributed between $\sim 5^\circ$ and $\sim 65^\circ$, resulting in a trigger threshold varying between
$\sim 100\ \mathrm{GeV}$ and $\sim 1\ \mathrm{TeV}$.

Data were analysed using two separate calibrations\cite{Aharonian:HESS_Calib} 
and analysis pipelines. The results presented here are based on the log-likelihood comparison of the
shower images with a precalculated semi-analytical model\cite{deNaurois:Model}. 

\section{Results}

\begin{figure}
\centering
  \includegraphics[width=0.45\textwidth]{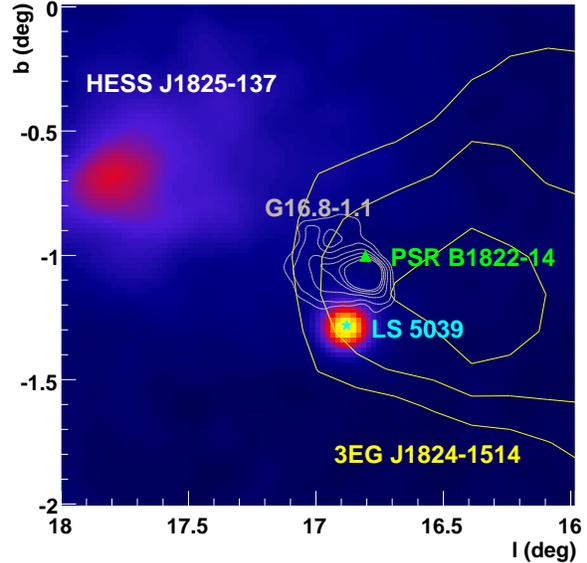}
\caption{H.E.S.S. excess sky map around LS~5039, smoothed by the instrument point spread function.
The blue star denotes the position of the VLBA source. The yellow contours correspond to the 68\%, 95\% and 
99\% confidence level region of the EGRET source 3EG J1824-1514. 
The extended source HESS J1825-137 observed in the same field of view can serve 
as a cross-check for timing-analysis.}
\label{fig:LSSkyMap}       
\end{figure}

After selection cuts, a total of 1969 $\gamma$-ray events were found within $0.1^\circ$ of the VLBA radio position\linebreak
of LS~5039\cite{Ribo:2002}, leading to a statistical significance of $40\sigma$ (Fig. \ref{fig:LSSkyMap}).
The best fit position is, in Galactic Coordinates, $l=16.879^\circ$, $b=-1.285^\circ$
with statistical and systematic uncertainties of respectively $\pm 12''$ and $\pm 20''$.
This position is compatible within $1\sigma$ with the VLBA position (denoted as a blue star in Fig. \ref{fig:LSSkyMap}) 
and with the Chandra source. We obtain an upper limit of $28''$ (at $1\sigma$)
on the VHE source extension.

\subsection{Timing Analysis}

The runwise VHE $\gamma$-ray flux at energies $\geq 1\ \mathrm{TeV}$ was decomposed into its frequency components
using the Lomb-Scargle periodogram\cite{Scargle:1982} (Fig. 1) which is appropriate for
unevenly sampled datasets such as those collected by H.E.S.S.

\begin{figure}
\centering
  \includegraphics[width=0.49\textwidth]{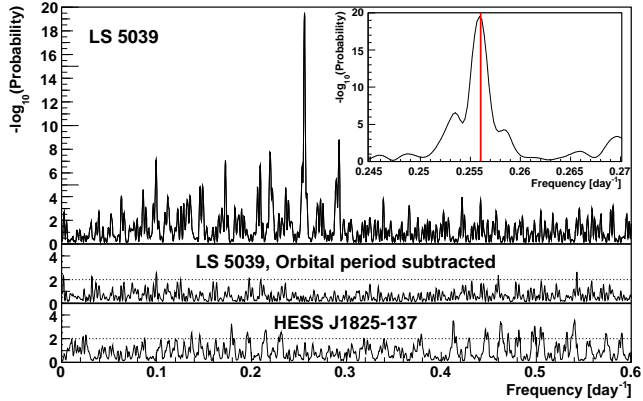}
\caption{Lomb-Scargle (LS) periodogram of the VHE runwise flux of LS~5039 above 1 TeV (Chance probability
to obtain the LS power vs. frequency).From \cite{Aharonian:HESS_LS5039:2}. Zoom: inset around
the highest peak, which corresponds to a period of $3.9078\pm 0.0015$ days. This period is found to 
be compatible with the orbital period determined by  Caseres {\it et al.}\cite{Casares:2005} and denoted 
as a red line on the inset. Middle: LS periodogram of the same data after subtraction of a pure 
sinusoidal component at the orbital period of 3.90603 days. The orbital peak is removed as expected, 
but also the satellite due to beat of the orbital period with various instrumental periods (see text).
Bottom: LS periodogram obtained on HESS J1825-137 using the same dataset (HESS J1825-137 is observed in the same field of view.)}
\label{fig:Lomb}       
\end{figure}

In order to reduce the effect of the varying instrument threshold, 
all events were used in the lightcurve determination and the runwise flux 
normalisation was extracted under the assumption of an average photon index 
derived from all data ($\Gamma=2.23$ for $dN/dU \propto E^{-\Gamma}$. The average
index assumption in this method notably increases the statistics and contributes 
only to a small  error on the derived flux above $1\ \mathrm{TeV}$). 

An obvious peak in the Lomb-Scargle periodogram occurs at the period $3.9078\pm 0.0015$ days,
consistent with the orbital period determined by Caseres {\it et al.}\cite{Casares:2005}
($3.90603 \pm 0.00017$) and excluding the earlier period of $4.4267\pm 0.0005$ days determined 
by McSwain et al\cite{McSwain:2001,McSwain:2004}. The peak is highly significant, with a chance
probability of $\sim 10^{-20}$ before trials (estimated via Monte-Carlo simulation of 
random fluxes time-series at the observations times as well as random shuffling of observed fluxes) and less 
than $10^{-15}$ after trials. Other peaks with chance probability less than $10^{-7}$-$10^{-8}$ are present
in the periodogram.

Fig \ref{fig:Lomb}, middle panel, shows the effect of subtracting a pure sinusoid at the orbital 
period. The orbital peak disappears as expected, but also the numerous satellite peaks, thus confirming
that these peaks are beat periods of the orbital period with the various gaps present in the H.E.S.S. dataset
(day-night cycle, moon period, annual period). The bottom panel of the same figure shows the result obtained
on the neighbouring source HESS J1825-137 observed in the same field of view as LS~5039. The HESS J1825-137
periodogram doesn't show any statistically significant peak, thus demonstrating that the observed
periodicity is genuinely associated with LS~5039.

\begin{figure}
\centering
  \includegraphics[width=0.45\textwidth]{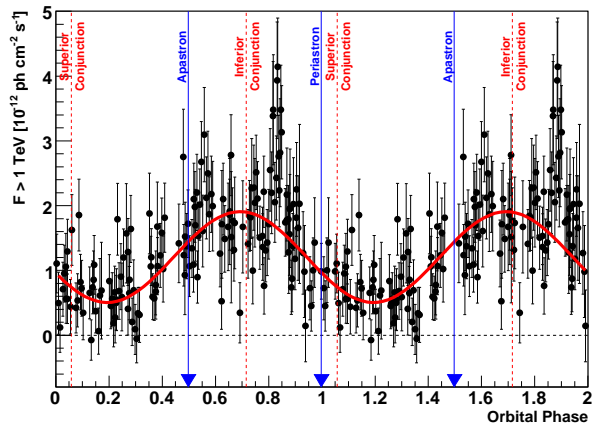}
\caption{Phasogram (Integral run-by-run $\gamma$-ray flux above 1 TeV as function of orbital phase) of LS~5039 
from H.E.S.S. data from 2004 to 2005, using the orbital ephemeris of\cite{Casares:2005}. Each run is $\sim 28$ minutes.
Two full phase periods are shown for clarity. The vertical blue arrows depict the respective phases of minimum ({\it periastron}) 
and maximum ({\it apastron}) binary separation. The vertical dashed red lines show the respective phases of inferior
and superior conjunction, when the star and the compact object are aligned along the observer's line of sight.
From \cite{Aharonian:HESS_LS5039:2}.}
\label{fig:Lightcurve}       
\end{figure}

\subsection{Flux Modulation}

The runwise Phasogram (Fig \ref{fig:Lightcurve}) of integral flux at energies $\geq 1\ \mathrm{TeV}$
vs. orbital phase ($\phi$) shows an almost sinusoidal behaviour, with the bulk of the
emission largely confined in a phase interval $\phi$ $\sim 0.45$ to $0.9$, covering about half
of the orbital period.
The thick red line in Fig \ref{fig:Lightcurve} represents the component at the orbital frequency
determined with the Lomb-Scargle coefficients. The emission maximum ($\phi\sim 0.7$) appear to lag behind 
the apastron epoch and to align better with the {\it inferior conjunction} ($\phi=0.716$), when the compact object lies
in front of the massive star (see Fig. \ref{fig:Cartoon}). The VHE flux minimum occurs at phase ($\phi \sim 0.2$), slightly further
along the orbit than {\it superior conjunction} ($\phi=0.058$).

Neither evidence for long-term secular variations in the VHE flux independent of the orbital modulation nor
any other modulation period are found in the presented H.E.S.S. data.

\subsection{Spectral Modulation}

Due to changing environment with orbital phase (magnetic field strength, stellar photon field, relative position
of compact object and star with respect to observer, \dots), the VHE $\gamma$-ray emission spectrum is expected 
to vary along the orbit. In such a binary system, the compact object environment can be modeled 
with a relatively good accuracy,  and the spectral modulation can therefore serve as a quite important diagnostic tool 
for disentangling the possible acceleration, cooling and absorption processes.

\begin{figure}
\centering
  \includegraphics[width=0.45\textwidth]{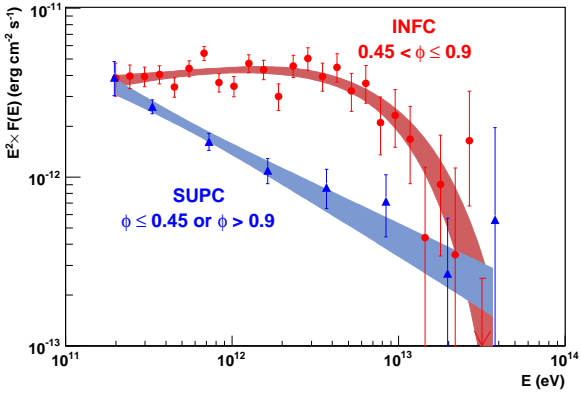}
\caption{Very high energy spectral energy distribution of LS~5039 for the two broad orbital phase
intervals defines in the text: {\bf INFC} $0.45 < \phi \leq 0.9$ (red circles) and its complementary {\bf SUPC} 
(blue triangles). The shades regions represent the $1\sigma$ confidence bands on the fitted functions.
Both spectral are mutually incompatible at the level of $\sim 2\times 10^{-6}$. A clear spectral hardening
is occurring in the $200\ \mathrm{GeV}$ to a few TeV range during the {\bf INFC} phase interval.
From \cite{Aharonian:HESS_LS5039:2}.}
\label{fig:SED}       
\end{figure}

We first define two broad phase interval: {\bf INFC} centered on the inferior conjunction ($0.45 < \phi \leq 0.9$)
and its complementary {\bf SUPC} centered on the superior conjunction, corresponding respectively to high
and low flux states.
Fig \ref{fig:SED} shows the VHE spectral energy distribution for these two phase intervals. The high state is consistent
with a hard power law with index $\Gamma= 1.85 \pm 0.06_{\mathrm{stat}} \pm 0.1_{\mathrm{syst}}$ with and exponential 
cutoff at $E_0 = 8.7 \pm 2.0\ \mathrm{TeV}$. In contrast, the spectrum for low state is compatible with a relatively steep
($\Gamma = 2.53 \pm 0.06_{\mathrm{stat}} \pm 0.1_{\mathrm{syst}}$) pure power law extending from $200\ \mathrm{GeV}$ to
$\sim 20\ \mathrm{TeV}$. The spectral shapes of these two states are mutually incompatible at the level of $\sim 2\times 10^{-6}$.
Interestingly, the flux appears to be almost unmodulated at $200\ \mathrm{GeV}$ as well as around $20\ \mathrm{TeV}$,
whereas the modulation is maximum around a few ($\sim 5$) TeV.

\begin{figure}
\centering
  \includegraphics[width=0.45\textwidth]{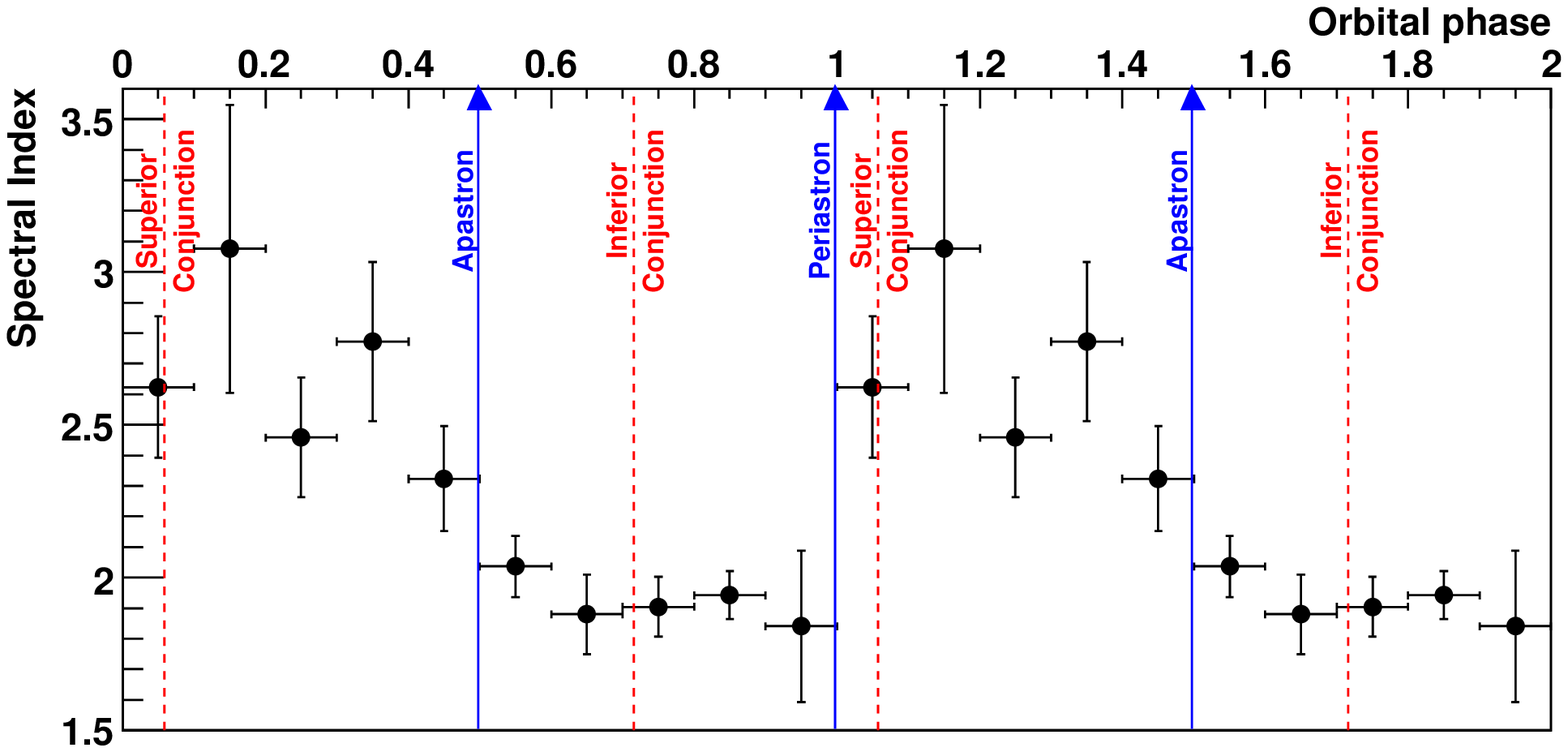}
  \includegraphics[width=0.45\textwidth]{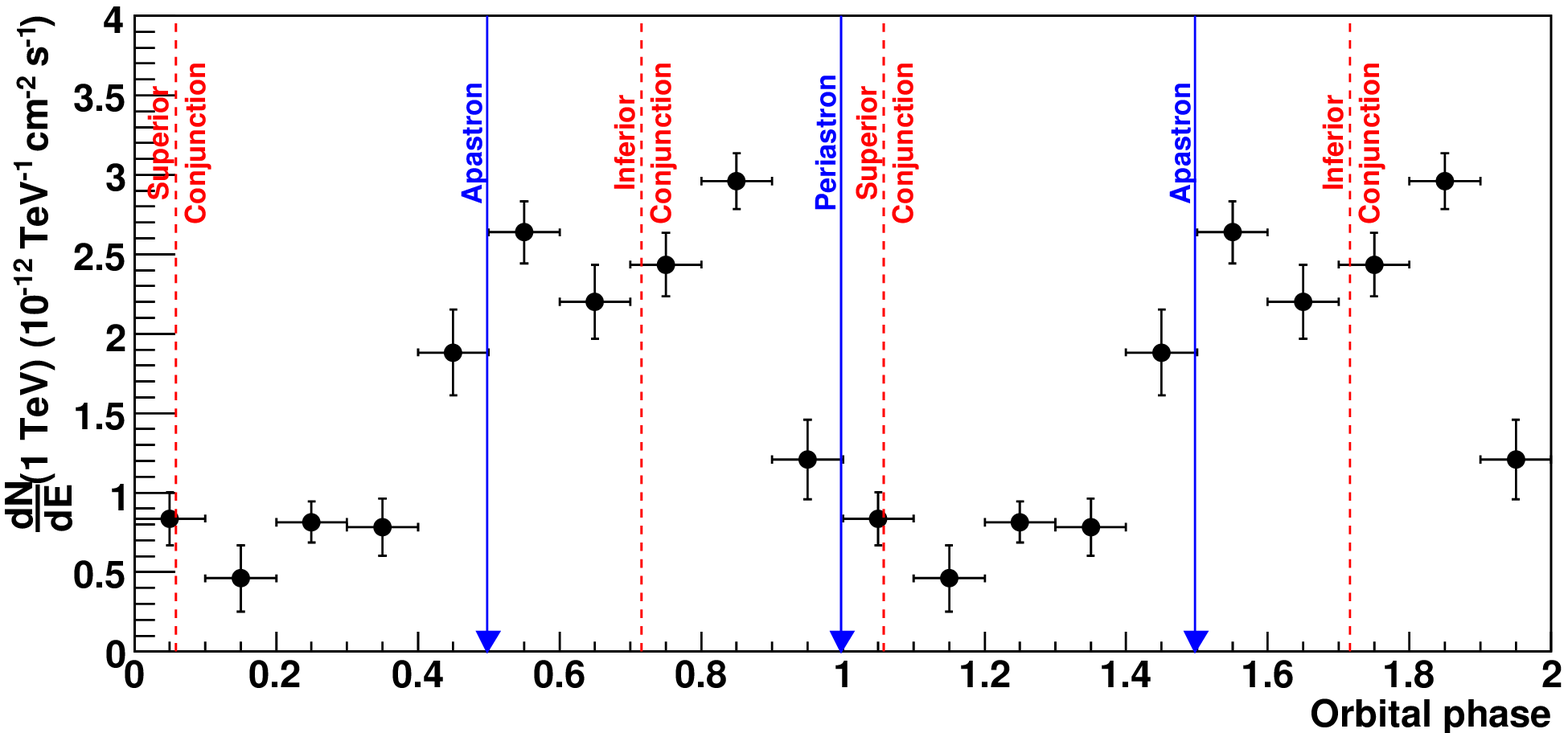}
\caption{Top: Fitted pure power-law photon index vs. phase interval of width $\Delta \phi = 0.1$.
Bottom: Differential flux at $1\, \mathrm{TeV}$ for the same phase interval. From \cite{Aharonian:HESS_LS5039:2}.}
\label{fig:PhaseSpectra}       
\end{figure}

When going to smaller phase interval ($0.1$ phase bins), the statistics at high energy becomes too low
to efficiently distinguish between power-law and more complicated shapes. Fig \ref{fig:PhaseSpectra}
shows the results (photon index and differential flux at $1\ \mathrm{TeV}$) of a pure power-law fit of 
the high energy spectra in $0.1$ orbital phase bins (restricted to energies below $5\ \mathrm{TeV}$ to avoid
systematic effect introduced by the high state cutoff).

\begin{figure}
\centering
  \includegraphics[width=0.45\textwidth]{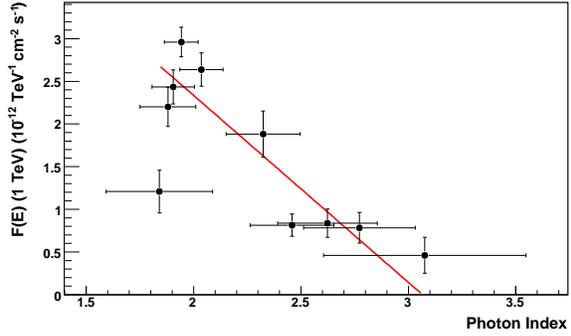}
\caption{Differential flux at $1\ \mathrm{TeV}$ vs. photon index.
The red line shows the best linear fit (correlation factor $r = 0.8$).}
\label{fig:Correlation}       
\end{figure}

The flux normalisation and photon index are strongly correlated, the flux being higher
when the spectrum is harder and vice-versa, as shown in Fig \ref{fig:Correlation}. 
The correlation factor is found to be $r \sim 0.8$. The photon index varies by more than
one unit along the orbit, whereas the flux normalization varies by a factor of more
than 5. Interestingly, a similar effect, however in a smaller variation range,
was found in X-rays\cite{Bosch:2005} where the photon index was found to vary between 1.7 and
2.2 with time. However, the X-ray phasogram exhibited a different picture than the VHE one,
with a flux maximum at $\phi \sim 0.2$  (close to the VHE flux minimum) and a second peak 
around $\phi\sim 0.8$ better aligned with the VHE flux.

\section{Interpretation}

\subsection{Gamma-Ray production}

The basic paradigm of VHE $\gamma$-ray production requires the presence of particles
accelerated to multi-TeV energies and a target comprising photons (for $\gamma$-ray production
through Inverse Compton effect) and/or matter of sufficient density (for $\gamma$-ray production
through pion decay in hadronic processes, e.g. \cite{Romero:2003}).
Several model classes are available to explain VHE emission from microquasars, 
differencing one from the other by the nature of accelerated particles and/or the location of the 
acceleration region. In jet-based models, particle acceleration could take place directly inside 
and along the jet, e.g. \cite[and references therein]{Bosch:2004}, and also in the jet termination 
shock regions\cite{Heinz:2002}. Non-jet scenarios are also available, e.g. \cite{Maraschi:1981,Dubus:2006b},
where the emission arises from the interaction of a pulsar wind with the stellar companion 
equatorial wind.

In LS~5039, the strong observed modulation provides new information about the physical processes
in microquasars, placing in particular strong constraints on the location of the acceleration region.

\subsection{Modulated absorption by pair creation}

\begin{figure}
\centering
  \includegraphics[width=0.4\textwidth]{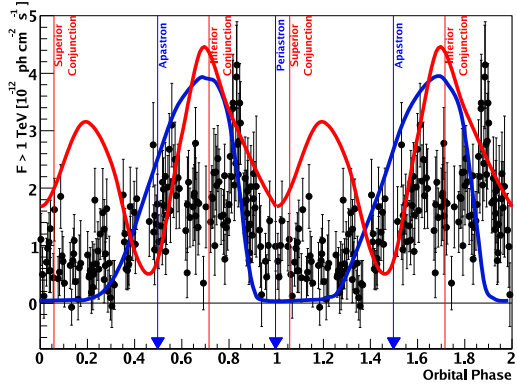}
\caption{Comparison of the observed phasogram (black points) with the expected flux modulation 
under a pure absorption scenario (blue line, adapted from \cite{Dubus:2006b})
and under a accretion disk-jet scenario (red line adapted from \cite{Paredes:2006}). 
In the accretion disk-jet scenario the modulation of the accretion rate of the stellar wind along the orbit is 
responsible for a modulation of the jet injection and therefore for a modulation of the particle acceleration in the jet.}
\label{fig:Prediction}       
\end{figure}

VHE $\gamma$-rays produced close enough to the stellar companion will
unavoidably suffer severe absorption via pair production ($e^+ e^-$) on its
intense photon field. Due to the angular dependence of the pair creation
cross-section and threshold, the optical depth will strongly depend
on the alignment between the $\gamma$-ray production region, the star and the
observer, leading to an orbital modulation of the VHE $\gamma$-ray flux
\cite{Protheroe:1987,Moskalenko:1995,Boettcher:2005,Dubus:2006a,Bednarek:2006}.
For the orbital geometry of LS~5039, a flux minimum is expected at the phase
of inferior conjunction, where the effect of the minimum absorption threshold 
and the maximum column density adds up.
The overall expected modulation for an emission close to the compact object
(Fig. \ref{fig:Prediction}, blue line adapted from \cite{Dubus:2006b})
agrees quite well with the observed picture, suggesting that absorption plays
an important role in the observed modulation.

\begin{figure}
\centering
 \includegraphics[width=0.4\textwidth]{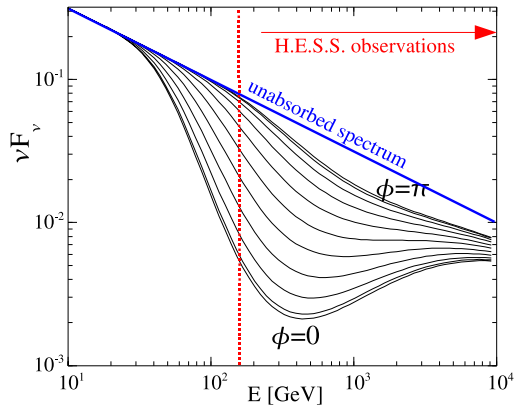}
\caption{Expected effect of the absorption on the $\gamma$-ray spectrum 
for different orbital phases. Adapted from \cite{Boettcher:2005}. The blue
line denotes the primary unabsorbed spectrum. The black lines show the
absorbed spectra for various orbital phases range between $0^\circ$ and $180^\circ$
(bottom to top). The sharply peaked pair production cross-section is expected to produce
a dip in the observed $\gamma$-spectrum just above the pair production 
production threshold, in the $200-500\ \mathrm{GeV}$.}
\label{fig:AbsorptionPrediction}       
\end{figure}

However, its has to be noted that a signal is observed by H.E.S.S. around phase $\phi=0$, 
which is unexpected under a pure absorption scenario. Detailed treatment of 
pair cascades, e.g. \cite{Bednarek:2006}, in which the VHE $\gamma$-ray energy is
reprocessed towards somewhat lower values, might help solving this issue. Another
key expectation from the absorption scenario is that the strongest absorption,
and hence modulation, should occur in the energy range $E\,\sim 0.2\ \mathrm{to}\ 2\ \mathrm{TeV}$
\cite{Dubus:2006a,Boettcher:2005} depending on orbital phase. Fig. \ref{fig:AbsorptionPrediction},
adapted from \cite{Boettcher:2005}, shows the expected evolution of the VHE spectrum with phase.
The flux dip at absorption maximum will produce a spectral softening at low 
energy ($\sim \leq 500 \mathrm{GeV}$) around phase $\phi\sim 0$, and a spectral hardening 
above. This picture is not consistent with the observed spectral modulation (Fig. \ref{fig:SED})
with notably an almost unmodulated flux at $\sim 0.2\ \mathrm{TeV}$, thus suggesting that additional
processes must be considered.

\subsection{Modulated particle acceleration}

VHE $\gamma$-ray production can be produced by accelerated electrons through the inverse-Compton (IC) 
scattering of stellar photons of the companion star, and/or accelerated hadrons through their interaction
with surrounding photons and particles. In this scenario, the efficiency of VHE $\gamma$-ray production will
peak around periastron, reflecting the higher target photon density.

The high temperature of the companion star means that IC $\gamma$-ray production proceeds primarily in 
the deep Klein-Nishina regime (where the IC cross-section is sharply reduced compared to the Thompson regime).
Due to changing magnetic field strength and target photon density along the orbit, the maximum electron 
energy will be increased by a factor of roughly 10 between periastron and apastron, where the IC cooling 
becomes less efficient (for a more detailed discussion, see \cite{Aharonian:HESS_LS5039:2}). 
This might agree quite well with the observed spectral modulation. Moreover,
synchrotron cooling takes over IC cooling above some energy $\epsilon \approx 6 [(B/{\rm G})(d/R_*)]^{-1}$
changing with orbital phase. Much stronger synchrotron losses\cite{Moderski:2005} would then produce a phase dependent 
spectral break, with a break energy lower by a factor of $\sim 10$ at periastron.

Other effects such as angular dependence of IC scattering\cite{Khang:1}
could also introduce spectral hardening at apastron phase.

\subsection{Modulated particle injection}

In an accretion disk-jet scenario, orbital modulation could arise
from modulation of the jet injection rate. Paredes et al\cite{Paredes:2006}
propose a leptonic jet model for LS~5039, where a slow equatorial stellar
wind induces a modulation of accretion rate (maximum shortly after periastron).
An additional stream wind formed next to periastron is required to reproduce
the doubled-peak X-ray lightcurve.

Their predicted VHE orbital modulation (Fig \ref{fig:Prediction}, red line)
does not agree well with the observed orbital modulation. In particular, the
enhancement in accretion rate shortly after periastron is not observed. However,
complication arise from the detail treatment of absorption effect, which is
this case depends on the assumption of the position of $\gamma$-ray 
production along the jet. In particular, emission in the small scale jets
would produce a similar absorption pattern as in Fig \ref{fig:Prediction}, blue line.

\section{Conclusion}

New observations by HESS have established orbital modulation of the VHE $\gamma$-ray flux 
and energy spectrum from the XRB LS~5039. The flux vs. orbital phase profile provides the 
first indication for $\gamma$-ray absorption within an astrophysical source, suggesting 
that a large part of the VHE $\gamma$-ray production region lies inside the pair absorption 
photosphere (within $\sim 1\,\mathrm{AU}$) around the massive stellar companion. However, not all of the observed 
effects can be explained by absorption alone. A detailed study is now required
to fully explain these new observations and understand the complex relationship 
between $\gamma$-ray absorption and production processes within these binary systems.

\begin{acknowledgements}
The support of the Namibian authorities and of the University of Namibia
in facilitating the construction and operation of H.E.S.S. is gratefully
acknowledged, as is the support by the German Ministry for Education and
Research (BMBF), the Max Planck Society, the French Ministry for Research,
the CNRS-IN2P3 and the Astroparticle Interdisciplinary Programme of the
CNRS, the U.K. Particle Physics and Astronomy Research Council\linebreak (PPARC),
the IPNP of the Charles University, the South African Department of
Science and Technology and National Research Foundation, and by the
University of Namibia. We appreciate the excellent work of the technical
support staff in Berlin, Durham, Hamburg, Heidelberg, Palaiseau, Paris,
Saclay, and in Namibia in the construction and operation of the
equipment.
\end{acknowledgements}




\end{document}